\documentclass[aps,prb,twocolumn,preprintnumbers,superscriptaddress]{revtex4}

\usepackage{graphicx}

\begin{document}

\draft

\title{Intersubband electronic Raman scattering in narrow GaAs single quantum wells dominated by single-particle excitations
\footnote{submitted to Phys. Rev. B.}
}

\author{Takeya Unuma}
\email[]{unuma@issp.u-tokyo.ac.jp}
\affiliation{
Institute for Solid State Physics, University of Tokyo, and CREST, JST, 
Kashiwa, Chiba 277-8581, Japan
}

\author{Kensuke Kobayashi}
\affiliation{
Institute for Solid State Physics, University of Tokyo, and CREST, JST, 
Kashiwa, Chiba 277-8581, Japan
}

\author{Aishi Yamamoto}
\affiliation{
Graduate School of Materials Science, Nara Institute of Science and Technology, Ikoma, Nara 630-0192, Japan
}

\author{Masahiro Yoshita}
\affiliation{
Institute for Solid State Physics, University of Tokyo, and CREST, JST, 
Kashiwa, Chiba 277-8581, Japan
}

\author{Yoshiaki Hashimoto}
\affiliation{
Institute for Solid State Physics, University of Tokyo, and CREST, JST, 
Kashiwa, Chiba 277-8581, Japan
}

\author{Shingo Katsumoto}
\affiliation{
Institute for Solid State Physics, University of Tokyo, and CREST, JST, 
Kashiwa, Chiba 277-8581, Japan
}

\author{Yasuhiro Iye}
\affiliation{
Institute for Solid State Physics, University of Tokyo, and CREST, JST, 
Kashiwa, Chiba 277-8581, Japan
}

\author{Yoshihiko Kanemitsu}
\altaffiliation[Also at: ]{Institute for Chemical Research, Kyoto University, 
Uji, Kyoto 611-0011, Japan.}
\affiliation{
Graduate School of Materials Science, Nara Institute of Science and Technology, Ikoma, Nara 630-0192, Japan
}

\author{Hidefumi Akiyama}
\affiliation{
Institute for Solid State Physics, University of Tokyo, and CREST, JST, 
Kashiwa, Chiba 277-8581, Japan
}

\date{8 June 2004}

\begin{abstract}
We measured resonant Raman scattering by intersubband electronic excitations in GaAs/AlAs single quantum wells (QWs) with well widths ranging from 8.5 to 18 nm. In narrow (less than $\sim 10$ nm) QWs with sufficiently high electron concentrations, only single-particle excitations (SPEs) were observed in intersubband Raman scattering, which was confirmed by the well-width dependence of Raman spectra. We found characteristic variations in Raman shift and line shape for SPEs with incident photon energy in the narrow QWs. 
\end{abstract}


\maketitle

\narrowtext
Intersubband electronic excitations in quantum wells (QWs) are strongly affected by many-body Coulomb interactions \cite{Ando:1982}. 
In electronic Raman scattering in doped QWs, two types of intersubband collective excitations have been confirmed by many researchers \cite{Pinczuk:1988} since the first reports by Abstreiter and Ploog \cite{Abstreiter:1979} and by Pinczuk \textit{et al.} \cite{Pinczuk:1979} in 1979:
a collective charge-density wave (CDW) appears if the polarizations of incident and scattered lights are parallel ($\parallel$), whereas a collective spin-density wave (SDW) appears if they are crossed ($\perp$). 

The CDW and SDW excitation energies are given by \cite{Ando:1982,note1}
\begin{eqnarray}
E_{\mathrm{CD}} &\!\! \approx \!\!& E_{10} + (\alpha - \beta)N_S, \\
E_{\mathrm{SD}} &\!\! \approx \!\!& E_{10} - \beta N_S
\end{eqnarray}
based on the local-density functional theory \cite{Kohn:1965,Hedin:1971,Bloss:1989}, where $N_S$ is the sheet electron concentration in a QW, $E_{10}$ is the intersubband energy separation that includes static many-body corrections, and $\alpha N_S$ and $\beta N_S$ are dynamical many-body corrections (called the depolarization shift and excitonic shift) due to the direct and exchange-correlation intersubband Coulomb interactions, respectively. 
In the crudest approximation \cite{note2,Bloss:1989}, the depolarization shift $\alpha N_S$ is proportional to $N_S d_{\mathrm{eff}}$ and the excitonic shift $\beta N_S$ is proportional to $(N_S/d_{\mathrm{eff}})^{1/3}$, where $d_{\mathrm{eff}}$ is the effective well width \cite{Ando:1982}. 
Experimentally, $\alpha N_S$ can be determined from the measured value of $E_{\mathrm{CD}} - E_{\mathrm{SD}}$; $\beta N_S$ can be estimated only when there is information about $E_{10}$. 

Later, in 1989, Pinczuk \textit{et al.} reported \cite{Pinczuk:1989} that not only CDW and SDW but also single-particle excitations (SPEs) are observed in intersubband electronic Raman scattering and their transition energy is $E_{10}$ for modulation-doped 25-nm GaAs/Al$_{0.3}$Ga$_{0.7}$As single QWs with $N_S = (1.5 - 3) \times 10^{11} \,\mathrm{cm^{-2}}$. 
SPEs are observed for both the parallel and crossed polarizations and seem to become stronger \cite{Pinczuk:1989,Peric:1993} at higher values of $N_S$. 

There are many other reports on intersubband electronic Raman scattering for wide (more than $\sim 20$ nm) GaAs QWs. 
It will be interesting to measure Raman scattering also in narrower QWs, considering $\beta/\alpha \sim (N_S d_{\mathrm{eff}}{}^2)^{-2/3}$. 
However, there are only a few reports for narrower GaAs QWs \cite{Abstreiter:1988,Ramsteiner:1990} because interface roughness makes Raman peaks broader and more difficult to observe. 
To our knowledge, such an experiment has never been reported for a single QW.

\begin{table*}
\caption{Main parameters of the samples. $L$ is the well width, $N_S$ is the sheet electron concentration, and $\mu$ is the mobility at 4.2 K. Theoretical values of the depolarization shift are calculated after Ref. 1.}
\begin{tabular}{cccccc}
\tableline
\tableline
Sample  &  Well/barrier  &  $L$ (nm)  &  $N_S\,(10^{12}\,\mathrm{cm}^{-2})$  &
$\mu\,(10^4\,\mathrm{cm^2/V s})$  
& Calculated depolarization shift (meV)\\
\tableline
N1 &  GaAs/AlAs  &  10  &  1.2  &  4.3  & 11 \\
N2  &  $\mathrm{In_{0.1}Ga_{0.9}As}$/AlAs 
                   &  8.5   &  2.0  &  2.8  & 14 \\
W &  GaAs/AlAs  &  18  &  0.63 &  44  & 9 \\
M &  GaAs/AlAs  &  13.5  &  1.1  &  5.6  & 12 \\
\tableline
\tableline
\end{tabular}
\label{samples}
\end{table*}

In this paper, we report resonant Raman scattering by the lowest ($0 \to 1$) intersubband electronic excitations in a set of modulation-doped GaAs/AlAs single QWs with well widths ranging from 8.5 to 18 nm. 
We found unexpected results that parallel- and crossed-polarization Raman spectra of narrow QWs had only a single peak at almost the same energy, though electron concentrations in the QWs were high enough that the theoretical values of the depolarization shift were more than 10 meV. 
On the other hand, spectra of a wide QW had a typical feature in that CDW, SDW, and SPE peaks all appeared clearly. 
In an intermediate-width QW we found no SDW peak, a very small CDW peak, and large SPE peaks, which led us to conclude that the peaks observed in narrow QWs correspond to the SPEs. 
Moreover, characteristic variations in Raman shift and line shape with incident photon energy were observed commonly in the narrow QWs. 
They were found to have a correlation with resonance strength and can be qualitatively understood by considering relaxation of the in-plane wave-vector conservation rule.

\begin{figure}[bt]
\begin{center}
\includegraphics[width=.35\textwidth]{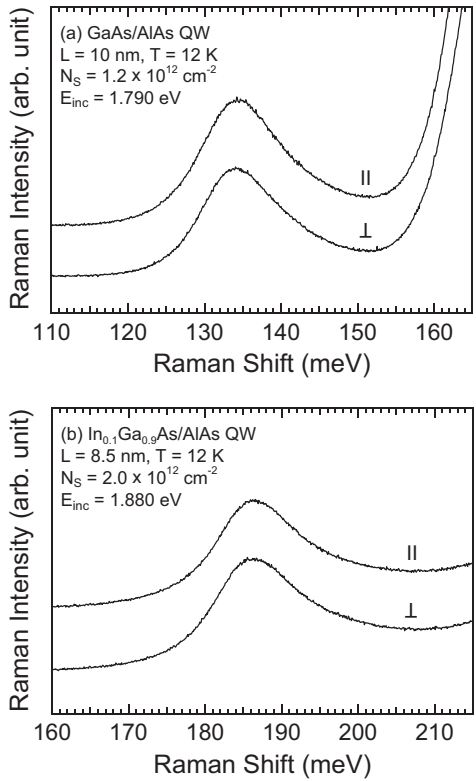}
\caption{Intersubband electronic Raman spectra of (a) 10-nm GaAs/AlAs (sample N1) and (b) 8.5-nm In$_{0.1}$Ga$_{0.9}$As/AlAs (sample N2) single QWs at 12 K. The $\parallel$ and $\perp$ marks mean that incident- and scattered-light polarizations were parallel and crossed, respectively.}
\label{narrow QWs}
\end{center}
\end{figure}

The samples used in this study were \textit{n}-type modulation-doped GaAs/AlAs single QWs grown by molecular beam epitaxy on (001) GaAs substrates: N1, N2, W, and M. 
Their well widths $L$ ranged from 8.5 to 18 nm, as shown in Table \ref{samples}. The QW structure of sample N1, for example, consisted of a Si-doped Al$_{0.33}$Ga$_{0.67}$As layer, a 4.0-nm undoped Al$_{0.33}$Ga$_{0.67}$As spacer layer, a 6.0-nm undoped AlAs barrier, a 10-nm undoped GaAs QW, a 6.0-nm undoped AlAs barrier, and a Si-doped Al$_{0.33}$Ga$_{0.67}$As layer. The other samples had similar QW structures. 
The samples were designed such that calculated values of the depolarization shifts \cite{Ando:1982} were more than 10 meV, which is larger than the linewidths of intersubband transition \cite{Unuma:2003} and large enough to see in Raman spectra. 
The electron concentrations were controlled mainly by the thickness of spacer layers. Exceptionally, $\mathrm{In}_{0.1} \mathrm{Ga}_{0.9} \mathrm{As}$ alloy was used for the QW layer of sample N2 in order to raise its electron concentration sufficiently. 
The sheet electron concentrations $N_S$ and mobilities $\mu$ at 4.2 K are also shown in Table \ref{samples}. 
The values of $N_S$ were confirmed by Shubnikov-de Haas measurements for samples N1, W, and M, and by Hall measurements for sample N2.

Raman spectra were measured at 12 K in a geometry of back-scattering normal to QW layers such that in-plane wave-vector transfer from an incident photon to an electron was less than $\sim 10^4 \,\mathrm{cm^{-1}}$. 
Typical values of incident laser power and spot size were 30 mW and 0.3 mm$^2$, respectively. 
Finding a suitable resonance condition is crucial for electronic Raman experiments in order to make Raman signals intensified and avoid their overlap with the strong photoluminescence (PL) from the GaAs QW or AlGaAs layers. 
 In our experiments, the incident photon energies $E_{\mathrm{inc}}$ were probably resonant with the energy gap between the electron first-excited ($E_1$) subband and the heavy-hole second-excited ($H_2$) subband in a GaAs QW \cite{Danan:1989}.
This was the only resonance condition that made intersubband Raman scattering observable in the energy range between the PL peak positions for the QW and AlGaAs layers in samples N1, N2, and M.

Figure \ref{narrow QWs} (a) shows typical intersubband Raman spectra of sample N1. The backgrounds of the spectra were due to PL from the QW.
The parallel- and crossed-polarization spectra have only a single peak at almost the same energies of 134.2 meV and 134.0 meV, respectively, while the theoretical value of the depolarization shift calculated after Ref. 1 was about 11 meV for the QW with $N_S = 1.2 \times 10^{12}\,\mathrm{cm}^{-2}$. 
A similar polarization dependence was observed in sample N2, as shown in Fig \ref{narrow QWs}. (b), where $N_S = 2.0 \times 10^{12}\,\mathrm{cm}^{-2}$ and the calculated depolarization shift was 14 meV. 
Apparently, these experimental results do not correspond to the conventional interpretation for wide QWs that CDW and SDW peaks are in the parallel- and crossed-polarization spectra, respectively, and their energy difference is the depolarization shift. 
We assigned the Raman peaks in Fig. \ref{narrow QWs} to SPEs, which was confirmed by the well-width dependence of Raman spectra as discussed in later paragraphs.

\begin{figure}[bt]
\begin{center}
\includegraphics[width=.42\textwidth]{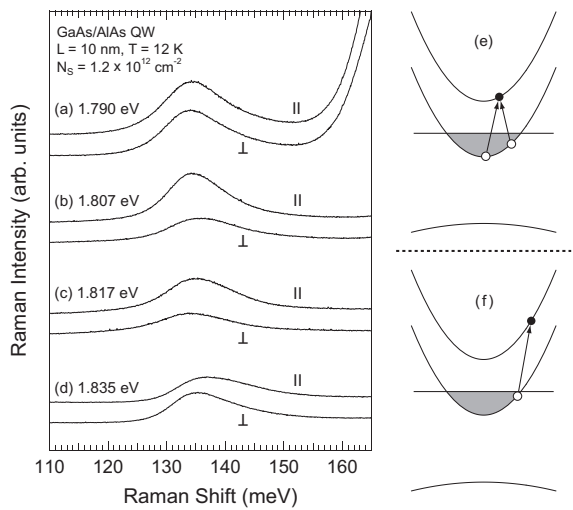}
\caption{Intersubband electronic Raman spectra of a 10-nm GaAs/AlAs QW (sample N1) at various incident photon energies $E_{\mathrm{inc}}$: (a) 1.790 eV, (b) 1.807 eV, (c) 1.817 eV, and (d) 1.835 eV.
$\parallel$: parallel polarizations, $\perp$: crossed polarizations. 
(e) and (f) are schematic diagrams of intersubband electronic excitations with the relaxation of the in-plane wave vector conservation rule at low and high $E_{\mathrm{inc}}$, respectively.}
\label{lambda-dep}
\end{center}
\end{figure}

The resonance condition for sample N1 had an allowable range of about 70 meV in the vicinity of $E_{\mathrm{inc}} = 1.80$ eV due to the electron Fermi distribution and the relaxation of the in-plane wave-vector conservation rule caused by interface roughness or other scatterers. 
In this range, we measured Raman spectra of sample N1 and found peculiar variations in Raman shift and line shape with incident photon energy. Similar characteristics were also found in sample N2. 

Figure \ref{lambda-dep} shows the spectra of sample N1 at various incident photon energies $E_{\mathrm{inc}}$: (a) 1.790 eV, (b) 1.807 eV, (c) 1.817 eV, and (d) 1.835 eV. 
In Fig. \ref{lambda-dep} (a), the parallel- and crossed-polarization spectra have almost the same peak positions. 
In Fig. \ref{lambda-dep} (b), however, the crossed-polarization spectrum has a slightly higher peak position than the parallel-polarization spectrum. 
In Fig. \ref{lambda-dep} (c), on the other hand, the parallel-polarization spectrum has a slightly higher peak position than the crossed-polarization spectrum. In Fig. \ref{lambda-dep} (d), the parallel-polarization spectrum again has a slightly higher peak position than the crossed-polarization spectrum, and their line shapes are very asymmetric. 
Note, on the whole in Fig. \ref{lambda-dep}, that the polarization dependence of the peak position at each $E_{\mathrm{inc}}$ is small compared with the theoretically expected depolarization shift of 11 meV, and that the Raman shift of the peak position for the parallel polarizations becomes larger as $E_{\mathrm{inc}}$ increases from 1.790 to 1.835 eV.

\begin{figure}[bt]
\begin{center}
\includegraphics[width=.35\textwidth]{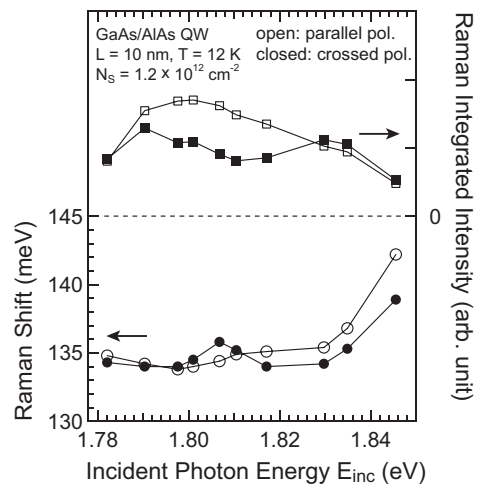}
\caption{Raman shift of peak position and Raman integrated intensity versus incident photon energy in a 10-nm GaAs/AlAs QW (sample N1). Open and closed symbols show the values for the parallel- and crossed-polarization spectra, respectively.}
\label{resonance curve}
\end{center}
\end{figure}

Figure \ref{resonance curve} shows the Raman shifts of peak position (circles) and the Raman integrated intensities (squares) versus incident photon energies in sample N1. Open and closed symbols correspond to the values for the parallel and crossed polarizations, respectively. 
In Fig. \ref{resonance curve}, there is a correlation between the Raman shift and the integrated intensity: the Raman shift becomes smaller as the resonance gets stronger. 
Near $E_{\mathrm{inc}} = 1.81$ eV, where the Raman shift is larger for the crossed polarizations than for the parallel polarizations, the resonance curve for the crossed polarizations has a dip. 
At $E_{\mathrm{inc}} = 1.83 - 1.85$ eV, where spectra have larger Raman shifts and very asymmetric line shapes as shown in Fig. \ref{lambda-dep} (d), the resonance is weaker for both the parallel and crossed polarizations.

\begin{figure}[bt]
\begin{center}
\includegraphics[width=.45\textwidth]{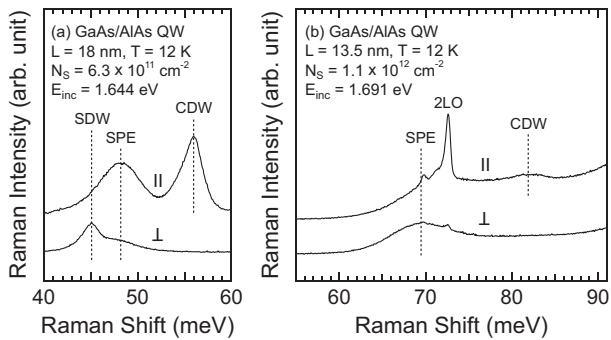}
\caption{Intersubband electronic Raman spectra of (a) 18-nm (sample W) and (b) 13.5-nm (sample M) GaAs/AlAs single QWs at 12 K. $\parallel$: parallel polarizations, $\perp$: crossed polarizations.}
\label{wide QWs}
\end{center}
\end{figure}

Next, to clarify the origin of the intersubband Raman peaks observed in samples N1 and N2, we also measured Raman spectra in wider QWs with well widths of 18 nm (sample W) and 13.5 nm (sample M) and investigated how spectral features changed with well width.
Figure \ref{wide QWs} (a) shows Raman spectra of sample W. The CDW peak at 56.1 meV and the SDW peak at 45.1 meV are clearly seen in parallel- and crossed-polarization spectra, respectively, as well as the SPE peaks at 48.3 meV in both the spectra. 
The energy difference of 11.0 meV between the CDW and SDW is in good agreement with the theoretical calculation of the depolarization shift.
This is a conventional result that follows Refs. 10, 11, and other reports.

Figure \ref{wide QWs} (b) shows Raman spectra of sample M, which had an intermediate well width: the large SPE peaks at 69.5 meV for both polarizations and the small CDW peak at 81.8 meV for the parallel polarizations. 
(Sharp peaks at 72.6 meV and 69.8 meV for the parallel polarizations are from 2-LO-phonon excitations in GaAs and AlGaAs layers, respectively.) 
The SDW peak is not seen, probably because it is too small or overlapped by the SPE peak. 
The change in spectral features with well width shows that the peaks observed in samples N1 and N2 should be assigned to the SPEs.

The electronic Raman process for intersubband SPEs is microscopically described as follows: An electron is excited from a valence band state to a conduction first-excited subband state $(\mathbf{k}_1, E_{10} + \varepsilon({\mathbf{k}_1}))$ by an incident photon with the resonant energy $E_{\mathrm{inc}}$, where $\varepsilon({\mathbf{k}}) = \hbar^2 \mathbf{k}^2/2m^*$ with $\mathbf{k}$ being the electron in-plane wave vector and $m^*$ being the effective mass.
Then, an electron in a conduction ground subband state $(\mathbf{k}_0, \varepsilon({\mathbf{k}_0}))$ recombines with the virtual hole remaining in the valence band state, which produces a scattered photon with energy $E_{\mathrm{scatt}}$. As the result, an intersubband SPE $(\mathbf{k}_0, \varepsilon({\mathbf{k}_0})) \to (\mathbf{k}_1, E_{10} + \varepsilon({\mathbf{k}_1}))$ occurs. 

In a geometry of back-scattering normal to QW layers, basically, the change in the in-plane wave vector $\mathbf{q} = \mathbf{k}_1 - \mathbf{k}_0$ is 0 and the Raman shift $E_{\mathrm{inc}} - E_{\mathrm{scatt}}$ for the intersubband SPE is $E_{10}$, since photon in-plane wave vectors are negligible. 
In fact, there are some dephasing mechanisms that cause broadening of the in-plane wave vectors of the electron and hole states involved, or the relaxation of in-plane wave-vector conservation rule; so $\mathbf{q} \neq 0$ intersubband SPE processes are also allowed.

Figure \ref{lambda-dep} (e) schematically shows a standard case of $|\mathbf{k}_1| < k_F$ for intersubband SPEs, where $k_F$ is the Fermi wave vector. 
In this case, various $\mathbf{q} \neq 0$ excitation processes only cause a symmetric width of Raman spectra around $E_{10}$, which we think corresponds to the case of $E_{\mathrm{inc}} < 1.83$ eV in our experiments. 
Near the higher limit of $E_{\mathrm{inc}}$ for intersubband Raman scattering, however, $|\mathbf{k}_1|$ is more than $ \sim k_F$ as shown in Fig. \ref{lambda-dep} (f). 
In this case, particular $\mathbf{q} \neq 0$ excitation processes such that $\mathbf{k}_0 \cdot \mathbf{q} > 0$ are dominant and $E_{\mathrm{inc}} - E_{\mathrm{scatt}} > E_{10}$. 
Thus, Raman spectra have a larger Raman shift and more asymmetric line shape (with a high-energy tail) at $E_{\mathrm{inc}} = 1.83 - 1.85$ eV than at $E_{\mathrm{inc}} < 1.83$ eV. 
It is not clear at present which carrier state was most deeply involved in the in-plane wave-vector conservation relaxation: the electron ground state, the electron first-excited state, or the virtual hole state.

Recently, Das Sarma and Wang have developed a theory of electronic Raman scattering explicitly considering the intermediate valence band state and the resonant excitation effect based on the random phase approximation (RPA) \cite{DasSarma:1999}, whereas standard RPA theories use a nonresonant approximation and neglect the intermediate valence band state. 
They have shown by numerical calculations that SPEs become strong only when the resonant excitation effect is included \cite{DasSarma:1999}. 
In our experiments the correlation between Raman shift and Raman integrated intensity with incident photon energy (Fig. \ref{resonance curve}) suggests that the resonant excitation effect is essential to the observed SPEs, and our results may be explained better by such an advanced theory. 

In summary, we have measured intersubband electronic Raman scattering in GaAs/AlAs single QWs with well widths ranging from 8.5 to 18 nm. 
The change in Raman spectral features with well width shows that only SPEs have been observed in narrow (8.5-nm and 10-nm) QWs for both parallel and crossed polarizations. 
Variations in Raman shift and line shape with incident photon energy have been found in the narrow QWs and qualitatively explained by considering $\mathbf{q} \neq 0$ intersubband SPEs inside and across $k_F$.

We are grateful to Professor T. Ando for his helpful discussions about the theory of intersubband electronic Raman scattering. 
This work was partly supported by a Grant-in-Aid from the Ministry of Education, Culture, Sports, Science and Technology, Japan. 
One of us (T.U.) also thanks the Japan Society for the Promotion of Science for partial financial support.


\begin{references}
\bibitem{note1}
We define $\alpha$ and $\beta$ following the convention of Ref. 10. They are related to Ando's convention by $\alpha = \alpha_{\mathrm{Ando}} E_{10}/(2N_S)$ and $\beta = \beta_{\mathrm{Ando}} E_{10}/(2N_S)$.
\bibitem{note2}
When $r_s = [4\pi a_B{}^3 n(z)/3]^{-1/3}$ is much less than unity, where $n(z)$ is the three-dimensional electron concentration and $a_B$ is the Bohr radius in GaAs, the exchange-correlation potential $V_{\mathrm{xc}}[n(z)]$ has an approximate dependence of $-1/r_s$, leading to $\beta N_S \sim (N_S/d_{\mathrm{eff}})^{1/3}$.
\end{references}
\end{document}